\documentclass[onecolumn]{aastex701}

\usepackage{amsmath}
\usepackage{tabularx}
\usepackage{longtable}
\usepackage{xcolor}
\usepackage{booktabs}

\begin{document}

\title{Multiband Color Monitoring of 3I/ATLAS through Ground-Based Relay Observations}
\shorttitle{Multiband Color Monitoring of 3I/ATLAS}
\author[0000-0001-9003-0999]{Ariel Graykowski}
\affiliation{SETI Institute, 339 Bernardo Ave, Suite 200, Mountain View, CA 94043, USA}
\email[show]{agraykowski@seti.org}
\correspondingauthor{Ariel Graykowski}

\author[0000-0002-4950-6323]{Bryce Bolin}
\affiliation{Eureka Scientific, Oakland, CA 94602, USA}
\email{bolin.astro@gmail.com}

\author[0009-0000-8709-5273]{Laura-May Abron}
\affiliation{Griffith Observatory, Los Angeles, CA, 90027, USA}
\email{lauramayabron@gmail.com}

\author[0000-0001-7016-7277]{Franck Marchis}
\affiliation{SETI Institute, 339 Bernardo Ave, Suite 200, Mountain View, CA 94043, USA}
\affiliation{SkyMapper Inc., San Francisco, CA, USA}
\email{fmarchis@seti.org}


\author[0000-0002-0967-0006]{Martin Ma\v{s}ek}
\affiliation{FZU - Institute of Physics of the Czech Academy of Sciences, Na Slovance 1999/2, CZ-182 21, Praha, Czech Republic}
\email{masekma@fzu.cz}

\author[0000-0002-5268-7735]{Filipp D. Romanov}
\affiliation{American Association of Variable Star Observers (AAVSO), 185 Alewife Brook Parkway, Suite 410, Cambridge, MA 02138, USA}
\email{filipp.romanov.27.04.1997@gmail.com}

\author[0000-0003-1169-4071]{Ahmed. M. Abdelaziz}
\affiliation{National Research Institute of Astronomy and Geophysics (NRIAG), 11421 Helwan, Cairo, Egypt}
\email{ahmed_astro84@nriag.sci.eg}

\author[0000-0002-2655-6459]{Dimitrios Athanasopoulos}
\affiliation{IAASARS, National Observatory of Athens, 15236 Penteli, Athens, Greece}
\email{d.athanaso1995@gmail.com}

\author[]{Matthew Belyakov}
\affiliation{Division of Geological and Planetary Sciences, California Institute of Technology, Pasadena, CA 91125, USA}
\email{mattbel@caltech.edu}

\author{Luca Buzzi}
\affiliation{Schiaparelli Astronomical Observatory, 21020 Varese VA, Italy}
\email{lucabuzzi.204@gmail.com}

% ...

\received{2026 September 17}
\accepted{---}
\submitjournal{AJ}

\begin{abstract}
We present multiband, long-baseline photometric observations of interstellar comet 3I throughout its 2025--2026 apparition using coordinated ground-based global relay observations. Our dataset combines measurements from professional observatories and citizen-operated Unistellar eVscopes distributed worldwide, providing dense temporal coverage from 2025 July 2 through 2026 April 1 and spanning the comet's pre- and post-perihelion trajectory. Broadband photometry was obtained in bandpasses equivalent to the Johnson--Cousins $B$ (436 nm), $V$ (545 nm), and $R$ (641 nm) filters and the Sloan $g$ (477 nm), $r$ (623 nm), and $i$ (763 nm) filters. The photometry was measured using projected aperture radii of approximately 10{,}000~km to provide a consistent probe of the inner coma across the heterogeneous dataset. We measure representative mean colors of $B-V=0.86\pm0.06$, $V-R=0.50\pm0.03$, $B-R=1.36\pm0.08$, and $g-r=0.58\pm0.08$, demonstrating a persistently red optical coma. Constant-color models provide an adequate description of the data, with little evidence for long-term color evolution with time or heliocentric distance despite substantial changes in the coma's brightness, gas production, and volatile composition. This suggests that the ensemble-averaged optical scattering properties of the coma remained relatively stable over the period sampled by our observations, even as other properties of the coma evolved. These observations provide the first densely sampled, apparition-long characterization of the broadband optical colors of an interstellar comet and establish a benchmark for comparison with future interstellar objects.
\end{abstract}


\section{Introduction}
\label{sec:intro}

\object{3I/ATLAS} (hereafter 3I) was discovered by the Asteroid Terrestrial-impact Last Alert System (ATLAS) on 2025 July 1 and was rapidly recognized as interstellar based on its strongly hyperbolic orbit \citep[e.g.,][]{Denneau2025,Bolin2025}. Early follow-up imaging resolved a compact coma, establishing 3I as an active comet and making it the second confirmed interstellar object, after 2I/Borisov, to exhibit clear cometary activity \citep[e.g.,][]{Bolin20202I,Bolin2020HST,JewittLuu2025activity}, in contrast with the asteroidal appearance of 1I/'Oumuamua \citep{Bolin20181I}. Prediscovery observations obtained by Rubin Observatory and the \textit{Transiting Exoplanet Survey Satellite} (\textit{TESS}) extended the observational record backward to 2025 May 7, when 3I was more than 6.3 au from the Sun \citep[e.g.,][]{Chandler2025,MartinezPalomera2025}. Backward orbital integrations and stellar encounter searches confirmed the interstellar origin of 3I but did not identify a unique parent star, indicating that its birth system remains unconstrained \citep{Guo20263I}.

Early optical and near-infrared observations showed that 3I possessed a red continuum broadly comparable to those of Solar System comets and other primitive small bodies \citep[e.g.,][]{Bolin2025,Belyakov2025RNAAS,Beniyama2025,Tonry2025}. Because 3I was active throughout these observations, its measured colors primarily trace sunlight scattered by material in the coma rather than reflection from the nucleus surface. Broadband colors therefore probe the ensemble-averaged optical properties of the coma, which depend on a combination of dust-particle size, composition, structure, and porosity \citep{ahearn1984,Kolokolova2024CometsIII}. They may also be affected by gas emission falling within the broad filter bandpasses \citep{Farnham2000}, especially at shorter wavelengths. Consequently, measurements of color evolution can trace evolution in the dust population and changes in the relative contributions of dust and gas. The persistent coma also complicates direct measurements of the nucleus: pre-perihelion \textit{Hubble Space Telescope} (\textit{HST}) observations placed an upper limit of $\sim2.8$ km on the effective nucleus radius \citep{Jewitt2025HST}, while post-perihelion coma-subtracted observations yielded $1.3\pm0.2$ km \citep{Hui2026}. Ground-based color measurements of 3I should therefore be interpreted primarily as measurements of coma material rather than of the nucleus surface.

Multiwavelength observations revealed substantial changes in the activity and coma composition of 3I through perihelion. Pre-perihelion James Webb Space Telescope (\textit{JWST})/NIRSpec spectroscopy and SPHEREx imaging spectrophotometry showed that the coma was strongly CO$_2$-dominated \citep{Cordiner2025,Lisse2025}. Dust production rates measured from visible broadband Nordic Optical Telescope (NOT) observations further supported this CO$_2$-dominated activity \citep{JewittLuu2025}. Optical spectroscopy traced increasing gas activity during the inbound leg, including the emergence and evolution of CN, Ni, C$_2$, and C$_3$ emission \citep{Bolin2025CN,SalazarManzano2025CN,Rahatgaonkar2025,Hoogendam2025}. Near perihelion, Jupiter Icy Moons Explorer (\textit{JUICE})/MAJIS and Solar and Heliospheric Observatory (\textit{SOHO})/SWAN observations revealed a dramatic increase in H$_2$O production \citep{bockelee2026,Combi2026}. By 2025 December, SPHEREx observations and \textit{JWST} spectroscopy revealed a shift toward CO-dominated activity \citep{Lisse2026Postperihelion,Roth2026Coma}. Dramatic post-perihelion changes in the volatile coma have been interpreted as evidence that 3I's exposed surface layers were substantially altered or removed, allowing more pristine subsurface material to contribute to the coma \citep{Belyakov2026MIRI,Belyakov2026Dust,zhao2026}.

Observations of the dust coma likewise indicate a complex particle population whose size, composition, and structure have been constrained using several complementary techniques. Early imaging provided only weak constraints on the particle-size distribution \citep{Bolin2025,SantanaRos2025}, while modeling of the coma morphology suggested that the optical scattering was dominated by relatively large, slowly moving particles of order 0.1~mm \citep{JewittLuu2025,Moreno26}. High-phase-angle imaging from Tianwen-1 similarly indicated particles hundreds of microns in size \citep{Ren2026}. \textit{JWST}/MIRI spectroscopy constrained the dust mineralogy through the 10~$\mu$m silicate feature, indicating predominantly amorphous silicates \citep{Belyakov2026Dust}. Polarimetry revealed an unusually deep and narrow negative polarization branch \citep{Gray2025}, with subsequent multiwavelength observations interpreted as evidence for refractory aggregates composed of submicron-scale monomers \citep{Choi2026}. Collectively, these observations demonstrate substantial complexity in both the volatile and dust components of the coma, motivating a search for corresponding changes in 3I’s broadband optical properties.

Numerous studies have measured broadband colors or optical continuum slopes of 3I, but most sample limited portions of the apparition and employ different filter systems, apertures, and instruments. Several studies reported persistently red colors over their respective observing intervals \citep{Rahatgaonkar2025,Gillan2026,Moreno26}, whereas ATLAS photometry revealed a very early transition from red to nearly solar $c-o$ color \citep{Tonry2025}. It therefore remains unclear whether the optical color of 3I evolved systematically through its passage through the inner Solar System or remained broadly stable despite the pronounced changes in its activity and coma composition. Addressing this question requires consistently analyzed, temporally dense color measurements spanning both the inbound and outbound portions of the apparition.

In this work, we present a densely sampled investigation of the broadband optical color evolution of 3I across its 2025--2026 apparition. We combine observations from the Unistellar Network with measurements from professional and citizen-operated telescopes distributed worldwide, providing extensive temporal coverage on both sides of perihelion. Using Johnson--Cousins $B-V$, $V-R$, and $B-R$ and Sloan $g-r$ colors mostly measured within projected apertures near 10{,}000~km, we test for color evolution with time and heliocentric distance and compare the coma colors before and after perihelion. Despite the substantial changes in the activity and composition of 3I, we find that the coma remained optically red with little evidence for systematic broadband color evolution over the portions of the apparition sampled by our observations.

\section{Observations}
\label{sec:obs}

The data used in this work were obtained through a global observing relay involving a variety of ground-based facilities worldwide, including professional observatories and citizen science observations obtained with Unistellar eVscopes through the Unistellar Network \citep[e.g.,][]{Graykowski2023Nature,Graykowski2025Hartley2}. Similar coordinated relay observations have been used for the follow-up and characterization of Solar System objects and astrophysical transients \citep[e.g.,][]{Hanus2018relay,Bolin2022Aylo,Perley2025}.

The resulting dataset is temporally dense but instrumentally heterogeneous. Observations were obtained using Johnson--Cousins $BVR$ and Sloan $gri$ filters, as well as several non-standard bandpasses that were transformed to standard photometric systems where appropriate. All observations, including their observing geometry and measured photometry, are summarized in Tables~\ref{tab:geometry} and \ref{tab:photometry}. Detailed descriptions of the contributing facilities, instrumentation, observing conditions, and reduction procedures are provided in Appendix~\ref{app:facilities}.

\section{Methods}
\subsection{Aperture photometry}
\label{sec:ap_phot}
For all datasets in this work, magnitudes were obtained using aperture photometry. To facilitate comparison among observations acquired across different geographic locations and with varying instrumentation, we adopted photometric apertures with projected radii as close as possible to 10{,}000 km at the comet. In practice, the exact aperture radius was chosen based on each image's pixel scale to best match the target physical scale. This choice provides a consistent measure of the inner-coma brightness across the heterogeneous dataset while minimizing systematic differences introduced by varying image sampling and observing geometry.

Most facilities in this work obtained observations through standard astronomical filters, including Johnson--Cousins $BVR$ and Sloan $gri$, and therefore directly reported filter-dependent magnitudes. Unistellar data, by contrast, required separating the Bayer channels and subsequently transforming them to estimate standard filter magnitudes. Prior to this transformation, the non-color-separated Unistellar images provide a filterless measurement broadly analogous to the Gaia $G$ bandpass. We therefore also examined the apparent Gaia $G$-band magnitude, $m_G$, derived from the non-color-separated Unistellar images. These measurements are tabulated in Table~\ref{tab:photometry}.

\begin{figure}[h]
    \centering
    \includegraphics[width=1\linewidth]{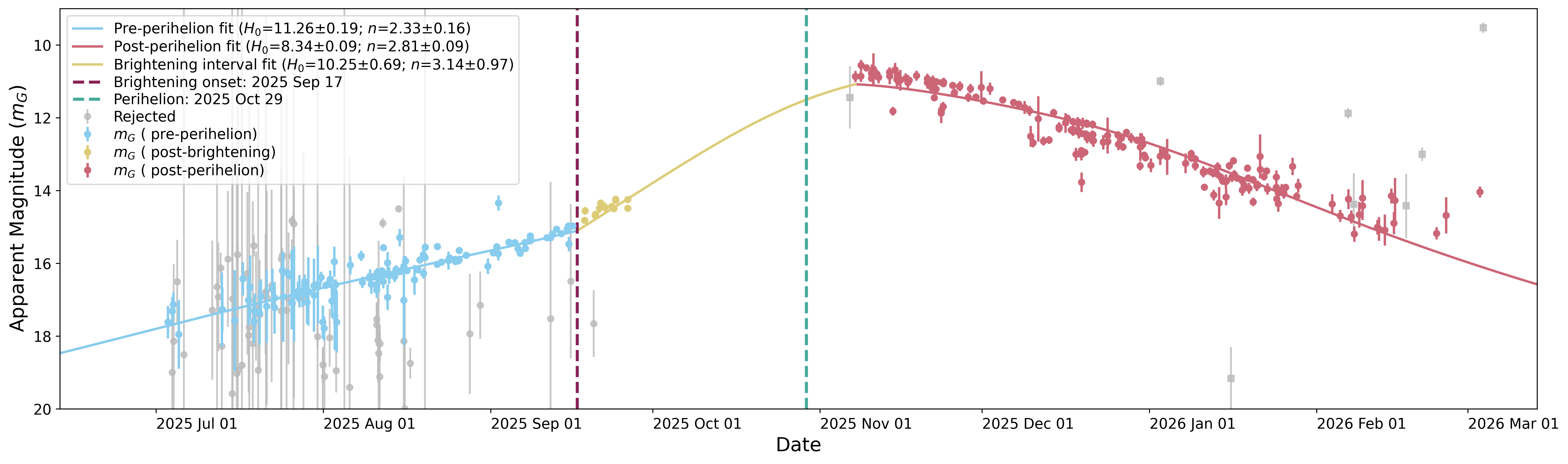}
    \caption{Secular light curve of 3I across its full observed apparition from Unistellar observations. Points show apparent Gaia $G$-band magnitudes ($m_G$) measured from non-color-separated images using apertures with projected radii near 10{,}000~km at the comet. Blue, yellow, and red circles denote observations in the pre-perihelion, brightening, and post-perihelion intervals, respectively, with corresponding fits of Equation~\ref{eq:comet_magnitude} shown as solid lines. Gray points were rejected by the quality cuts described in the text. Vertical dashed lines mark the onset of the brightening interval on 2025 September 17 and perihelion on 2025 October 29. No phase-angle correction is applied; the effect of the Schleicher--Marcus comet-dust phase function is discussed in Section~\ref{sec:ap_phot}.}
    \label{fig:lc}
\end{figure}

Prior to fitting, observations were subject to two quality cuts applied identically to both the pre- and post-perihelion datasets. First, points with photometric uncertainties exceeding five times the median uncertainty of the respective dataset were rejected as unreliable measurements. Second, an initial fit of Equation~\ref{eq:comet_magnitude} was made to the surviving low-uncertainty observations, and points whose residuals from that preliminary fit exceeded three times the robust scatter, estimated using the normalized median absolute deviation, were rejected as outliers. Rejected observations are shown as gray points in Figure~\ref{fig:lc}.

To characterize the brightness evolution, we fit a standard cometary magnitude law of the form
\begin{equation}
\label{eq:comet_magnitude}
m_G = H_0 + 5 \log_{10}(\Delta) + 2.5n \log_{10}(r_h),
\end{equation}
where $H_0$ is a fitted constant, $n$ is the heliocentric brightening parameter, and $\Delta$ and $r_h$ are the geocentric and heliocentric distances, respectively, in au. The comet exhibited a distinct brightening episode beginning around 2025 September 17, approximately six weeks before perihelion, during which it departed markedly from the preceding secular trend. We therefore fit Equation~\ref{eq:comet_magnitude} separately to the pre-perihelion, brightening, and post-perihelion intervals.

The pre-perihelion fit yields $H_0=11.26\pm0.19$ and $n=2.33\pm0.16$. During the brightening interval, we obtain $H_0=10.25\pm0.69$ and $n=3.14\pm0.97$. The larger uncertainties in this interval reflect its shorter temporal baseline and the smaller number of observations available before the comet became unobservable near perihelion. The post-perihelion observations were fit independently, yielding $H_0=8.34\pm0.09$ and $n=2.81\pm0.09$. The three fitted intervals are shown in Figure~\ref{fig:lc}.

The photometry shown in Figure~\ref{fig:lc} has not been corrected for solar phase angle. Our observations span $\alpha=0.9^\circ$--$30.6^\circ$, corresponding to approximately 0.9~mag of variation in dust brightness under the standard Schleicher--Marcus composite comet-dust phase function \citep{marcus2007,schleicher2011}. To assess the potential influence of phase angle, we repeated the analysis after correcting each measurement to $\alpha=0^\circ$ using this phase function. The phase-corrected fits yield $H_0=9.65\pm0.19$ and $n=3.14\pm0.16$ for the pre-perihelion interval, $H_0=9.95\pm0.69$ and $n=2.49\pm0.98$ for the brightening interval, and $H_0=7.04\pm0.09$ and $n=3.58\pm0.09$ for the post-perihelion interval.

We retain the uncorrected photometry as our nominal representation because the Schleicher--Marcus function describes the average dust-scattering behavior of Solar System comets and may not be appropriate for 3I \citep{zhang2026}. Polarimetric observations indicate unusual scattering behavior from 3I's dust, suggesting grain properties that may differ from those underlying standard cometary phase functions \citep{Gray2025,Choi2026}. We therefore use the phase-corrected fits only to illustrate the potential influence of phase angle. We do not apply phase corrections to the multiband photometry in the following sections because, without a wavelength-dependent phase function for 3I, the same correction would apply to each band and cancel in the color indices.

\subsection{Unistellar color extraction and transformation}
\label{sec:unistellar_color}

Because Unistellar instruments do not use standard astronomical filters, their color measurements require an additional calibration step before they can be directly compared with filtered observations from other facilities. For each image, we separated the Bayer matrix into its instrumental blue, green, and red channels and performed aperture photometry on each channel independently. We then transformed these instrumental Unistellar magnitudes, $B_{\mathrm{uni}}$, $G_{\mathrm{uni}}$, and $R_{\mathrm{uni}}$, to Johnson--Cousins $BVR$ using the color-transformation equations derived by \citet{Perrocheau2025}. The resulting magnitudes are tabulated in Table~\ref{tab:photometry}. Although our photometric calibration relied on various reference stars that spanned a range of spectral classes, the linear transformation equations of \citet{Perrocheau2025} remain robust, and any systematic residuals resulting from the red color of 3I are well within our quoted photometric uncertainties.

This procedure was performed on individual exposures rather than on the final co-added images from each Unistellar observation. After co-addition, the Bayer pattern is no longer preserved in a form that allows straightforward separation of the blue, green, and red channels. Performing the analysis on individual exposures reduces the effective signal in each channel relative to the full, non-color-separated image by approximately a factor of 2 for the green channel and 4 for the blue and red channels, thereby increasing the scatter and uncertainty of the transformed $B$, $V$, and $R$ measurements. We therefore determined the temporal interval over which the Unistellar color-separated photometry remained reliable directly from the observed photometric scatter rather than adopting a limiting magnitude.

\begin{figure}[h]
    \centering
    \includegraphics[width=1\linewidth]{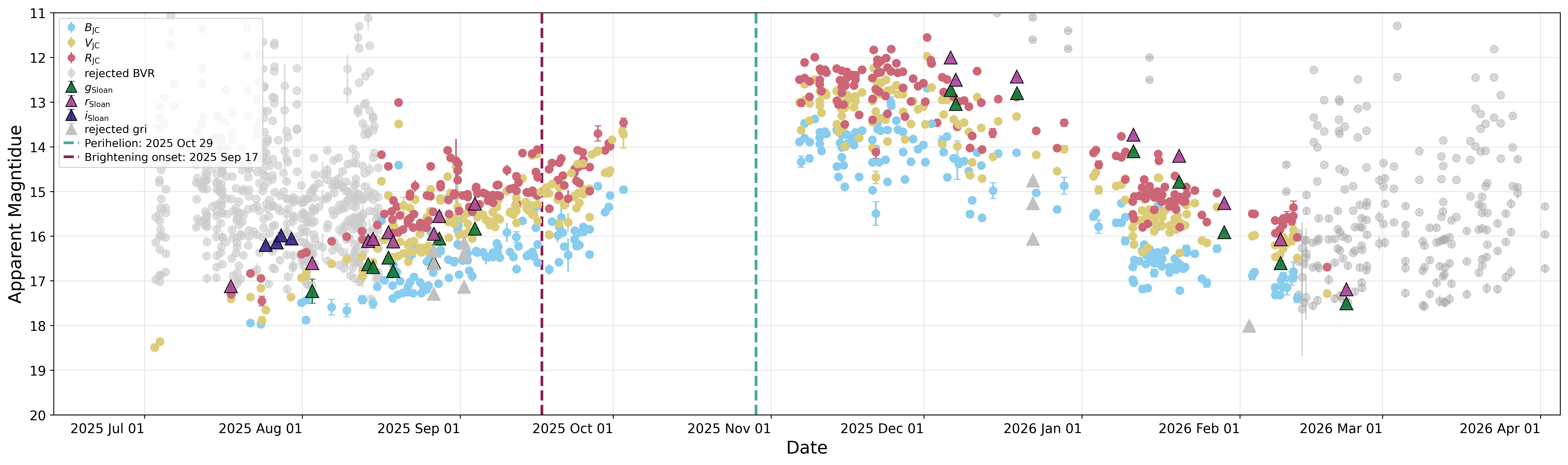}
    \caption{Apparent magnitudes of 3I as a function of time. Accepted Johnson--Cousins $B$, $V$, and $R$ measurements are shown as blue, yellow, and red circles, respectively, while external Sloan $g$, $r$, and $i$ measurements are shown as green, magenta, and purple triangles, respectively. Rejected Johnson--Cousins $BVR$ and Sloan $gri$ measurements are shown in gray. The vertical dashed lines mark the onset of the pre-perihelion brightening on 2025 September 17 and perihelion on 2025 October 29.}

    \label{fig:BVR_lc}
\end{figure}

To determine the temporal range over which the Unistellar color-separated photometry remained reliable, we quantified the scatter in the $B$, $V$, and $R$ measurements as a function of time. Because the brightness of 3I changes throughout the apparition, a local linear trend was independently removed from each band within rolling windows of 15 Unistellar observations, and the scatter of the residuals was measured using the median absolute deviation (MAD). For each window, we adopted the largest scatter among the three bands, $S=\max(\sigma_B,\sigma_V,\sigma_R)$, such that the least precise color channel determined the photometric quality. We then defined baseline intervals representative of reliable color-separated photometry and used the distribution of $S$ within each baseline to establish its normal scatter and variability. Pre-perihelion, the baseline was selected from the low-scatter portion of the data from 2025 August 31 to September 15, yielding a median $S=0.32$ mag and a $3\sigma$ upper threshold of $S=0.71$ mag. Searching backward from this reliable interval, we identify 2025 August 16 as the point before which the scatter exceeds this threshold for at least three consecutive rolling windows. Post-perihelion, when 3I remained sufficiently bright for a longer period, we used the first 60 days of the post-perihelion observations as the baseline. This yields a median $S=0.34$ mag and a $3\sigma$ upper threshold of $S=0.74$ mag. The scatter exceeds this threshold for at least three consecutive rolling windows beginning on 2026 February 12. We therefore restrict the Unistellar $B$, $V$, and $R$ analysis to observations obtained between 2025 August 16 and 2026 February 12. Measurements outside this interval are shown in gray in Figure~\ref{fig:BVR_lc}.

The combined Johnson--Cousins $B$, $V$, and $R$ measurements, together with the Sloan $g$, $r$, and $i$ measurements, are shown in Figure~\ref{fig:BVR_lc}. A few Sloan $g$, $r$, and $i$ measurements were rejected because they were anomalously faint relative to the surrounding observations. 

\section{Results}
\label{sec:results}

\begin{figure}[h]
    \centering
    \includegraphics[width=1\linewidth]{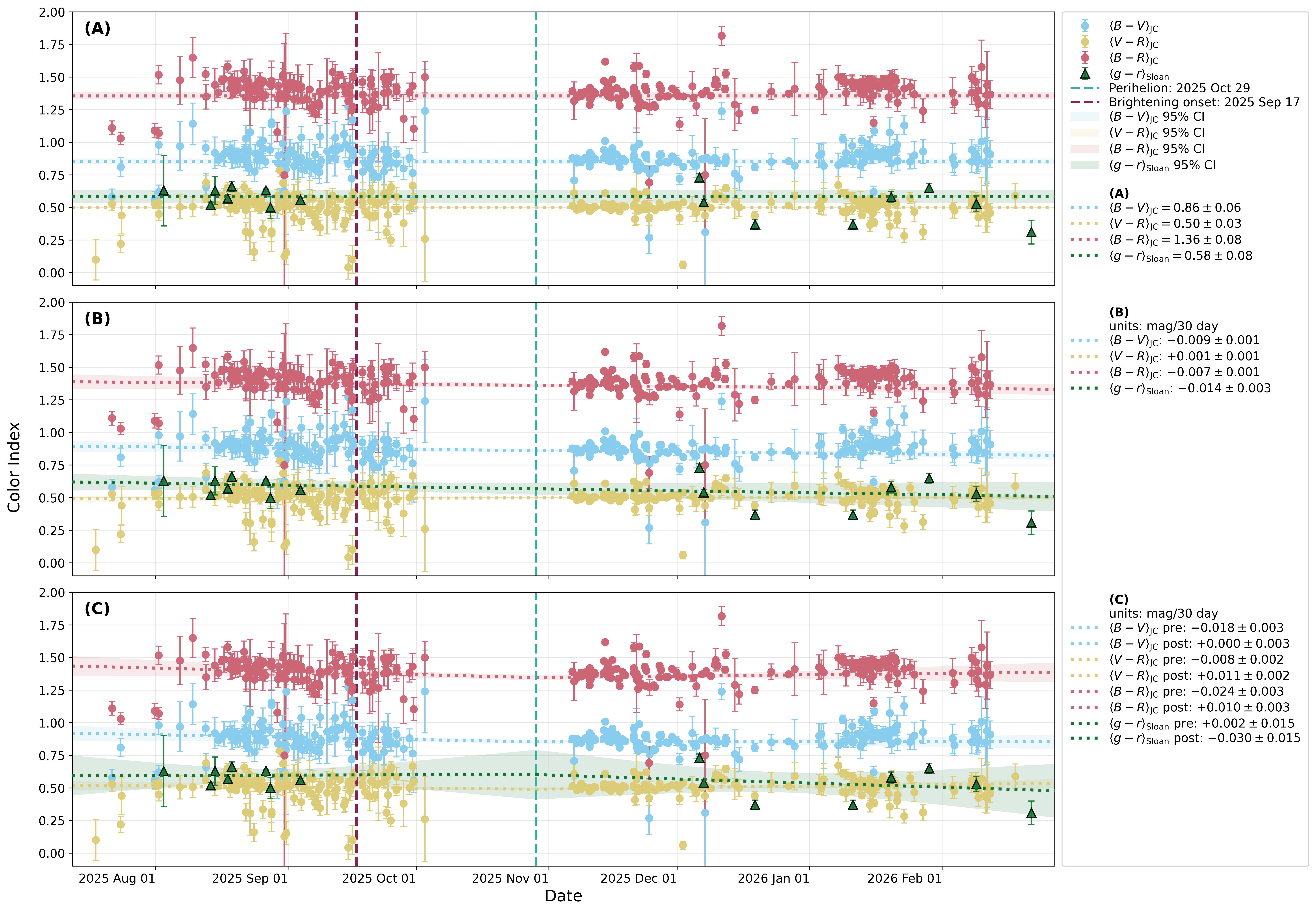}
    \caption{Temporal evolution of the Johnson--Cousins $B-V$, $V-R$, and $B-R$ and Sloan $g-r$ color indices of 3I. Johnson--Cousins $B-V$, $V-R$, and $B-R$ measurements are shown as circles, while Sloan $g-r$ measurements are shown as triangles. The vertical dashed lines mark the onset of the pre-perihelion brightening on 2025 September 17 and perihelion on 2025 October 29. Dotted lines show the best-fitting (A) constant, (B) linear, and (C) piecewise-linear models, with the latter allowing independent slopes before and after perihelion. Shaded regions show the scaled 95\% confidence intervals (CIs) on the fitted models (note: these are not prediction intervals), accounting for the elevated reduced $\chi^2$ of each color index. Fit parameters and their formal uncertainties are given in the legend.}
    \label{fig:colors}
\end{figure}

We plot the Johnson--Cousins $B-V$, $V-R$, and $B-R$ colors together with the Sloan $g-r$ colors in Figure~\ref{fig:colors}. The weighted-mean colors and their standard deviations are $B-V=0.86\pm0.06$, $V-R =0.50\pm0.03$, $B-R=1.36\pm0.08$, and $g-r=0.58\pm0.08$. We use the standard deviation of the measurements to characterize the uncertainty on these representative colors because the constant fits have large reduced chi-squared values ($\chi^2_\nu = 11.2$, 8.3, 19.9, and 10.6 for $B-V$, $V-R$, $B-R$, and $g-r$, respectively), indicating that the observed scatter substantially exceeds that expected from the formal photometric uncertainties alone. Inter-observatory systematic offsets likely contribute to this excess scatter. Although some observations included both Sloan $r$ and $i$ measurements, none of the resulting $r-i$ colors passed our quality cuts, so we do not report a representative $r-i$ color.

We adopt the constant-color model as our baseline and test for temporal color evolution by fitting two additional models to each index: a linear slope in time centered on perihelion and a piecewise-linear model with independent slopes before and after perihelion (Table~\ref{tab:color_fits}). The fitted linear slopes have magnitudes of at most $0.015$~mag per 30~days, while the piecewise pre- and post-perihelion slopes have magnitudes of at most $0.025$ and $0.035$~mag per 30~days, respectively. The direction of the fitted evolution is not consistent among the four color indices and the values of $\chi^2_\nu$ do not show a consistent decrease with increasing model complexity.

{\setlength{\tabcolsep}{3pt}
\begin{deluxetable*}{lcccccccccc}
\tabletypesize{\footnotesize}
\tablecaption{Color model fits for 3I.\label{tab:color_fits}}
\tablewidth{0pt}
\tablehead{
\colhead{Index} & \colhead{$N$} &
\colhead{Mean} & \colhead{$\chi^2_{\nu,\rm C}$} &
\colhead{Slope\tablenotemark{a}} & \colhead{$\chi^2_{\nu,\rm L}$} & \colhead{$\Delta\chi^2_\nu$\tablenotemark{b}} &
\colhead{Pre\tablenotemark{a}} & \colhead{Post\tablenotemark{a}} & \colhead{$\chi^2_{\nu,\rm PW}$} & \colhead{$\Delta\chi^2_\nu$\tablenotemark{c}}
}
\startdata
$B-V$ & 254 & $0.86\pm0.01$  & 11.2 & $-0.009\pm0.001$ & 11.1 & $-0.1$ & $-0.018\pm0.003$ & $+0.000\pm0.003$ & 11.1 & $+0.0$ \\
$V-R$ & 264 & $0.50\pm0.003$ & ~8.3 & $+0.001\pm0.001$ & ~8.4 & $+0.0$ & $-0.008\pm0.002$ & $+0.011\pm0.002$ & ~8.2 & $-0.2$ \\
$B-R$ & 254 & $1.36\pm0.01$  & 19.9 & $-0.007\pm0.001$ & 19.8 & $-0.1$ & $-0.024\pm0.003$ & $+0.010\pm0.003$ & 19.7 & $-0.1$ \\
$g-r$ &  15 & $0.58\pm0.03$  & 10.6 & $-0.015\pm0.004$ & 10.1 & $-0.5$ & $+0.004\pm0.015$ & $-0.034\pm0.016$ & 10.8 & $+0.7$ \\
\enddata
\tablenotetext{a}{Slopes (Pre and Post) in mag per 30~days; formal fitting errors only (see text). Pre/Post are the piecewise-linear slopes before and after perihelion; the PW model is fitted to all data simultaneously and yields a single $\chi^2_{\nu,\rm PW}$, not separate values per segment.}
\tablenotetext{b}{Change in $\chi^2_\nu$ from constant (C) to linear (L). Negative = improvement.}
\tablenotetext{c}{Change in $\chi^2_\nu$ from linear (L) to piecewise (PW). Negative = improvement.}
\end{deluxetable*}}

The parameter uncertainties reported in Table~\ref{tab:color_fits} are formal fitting errors. Because $\chi^2_\nu \gg 1$, these formal uncertainties underestimate the scatter represented by the data. We therefore scale the parameter uncertainties by $\sqrt{\chi^2_\nu}$, corresponding to factors of 3.3, 2.9, 4.5, and 3.2 for $B-V$, $V-R$, $B-R$, and $g-r$, respectively. The resulting scaled 95\% CIs for the constant colors and for the linear and piecewise-linear slopes are listed in Table~\ref{tab:color_fit_ci}. The corresponding confidence bands are shown as the shaded regions in Figure~\ref{fig:colors}. 

{\setlength{\tabcolsep}{6pt}
\begin{deluxetable*}{lcccc}
\tabletypesize{\footnotesize}
\tablecaption{Scaled 95\% confidence intervals for the color-model fits of 3I.\label{tab:color_fit_ci}}
\tablewidth{0pt}
\tablehead{
\colhead{Index} &
\colhead{Constant\tablenotemark{a}} &
\colhead{Linear slope\tablenotemark{b}} &
\colhead{Pre-perihelion slope\tablenotemark{b}} &
\colhead{Post-perihelion slope\tablenotemark{b}}
}
\startdata
$B-V$ & $[0.835,\ 0.875]$ & $[-0.019,\ +0.000]$ & $[-0.037,\ +0.001]$ & $[-0.019,\ +0.019]$ \\
$V-R$ & $[0.493,\ 0.505]$ & $[-0.004,\ +0.007]$ & $[-0.019,\ +0.003]$ & $[+0.000,\ +0.022]$ \\
$B-R$ & $[1.331,\ 1.382]$ & $[-0.020,\ +0.005]$ & $[-0.050,\ +0.002]$ & $[-0.016,\ +0.036]$ \\
$g-r$ & $[0.532,\ 0.637]$ & $[-0.037,\ +0.008]$ & $[-0.090,\ +0.098]$ & $[-0.134,\ +0.066]$ \\
\enddata
\tablenotetext{a}{95\% confidence interval on the fitted constant color, in mag. The uncertainty on the constant fit is taken as the larger of the formal uncertainty scaled by $\sqrt{\chi^2_\nu}$ and the uncertainty estimated by resampling observing dates, which accounts for correlations among measurements obtained on the same UTC date.}
\tablenotetext{b}{95\% confidence interval on the fitted slope, in mag per 30~days. Slope confidence intervals are calculated after scaling the formal fit covariance by the reduced $\chi^2$ of the corresponding model.}
\end{deluxetable*}}

We compare the representative colors measured in this study with previous measurements in Table~\ref{tab:color_comparison}. Our values are somewhat less red than the near-discovery measurements, which were obtained primarily before the time range covered by our color measurements, but are broadly consistent with the range of broadband colors reported by other studies. Independent spectroscopic studies likewise show a red optical continuum, although the reported spectral slopes vary with wavelength and observing epoch \citep[e.g.,][]{AlvarezCandal2025,DeLaFuenteMarcos2025,Yang2025,Rahatgaonkar2025,Hoogendam2025,Kawakita2026,Medler2026}.

\begin{deluxetable*}{lccccc}
\tablecaption{Comparison of average broadband color measurements of 3I. \label{tab:color_comparison}}
\tablewidth{0pt}
\tablehead{
\colhead{Study} & \colhead{Date range} & \colhead{$B-V$} & \colhead{$V-R$} & \colhead{$g-r$} & \colhead{$r-i$}
}
\startdata
\textbf{This work} & 2025 Jul 17--2026 Feb 21 & $0.86 \pm 0.06$ & $0.50 \pm 0.03$ & $0.58 \pm 0.08$ & \nodata \\
\citet{Chandler2025} & 2025 Jun 21--Jul 20 & \nodata & \nodata & $0.657 \pm 0.013$ & $0.235 \pm 0.018$ \\
\citet{Bolin2025} & 2025 Jul 2--6 & $0.98 \pm 0.23$ & $0.71 \pm 0.09$ & $0.84 \pm 0.05$ & $0.16 \pm 0.03$ \\
\citet{Seligman2025} & 2025 Jul 2--3 & \nodata & \nodata & $0.85 \pm 0.03$ & $0.25 \pm 0.03$ \\
\citet{Opitom2025} & 2025 Jul 3 & $1.12 \pm 0.14$ & $0.57 \pm 0.09$ & \nodata & \nodata \\
\citet{Bolin2025JulAtel} & 2025 Jul 6 & \nodata & \nodata & $0.81 \pm 0.06$ & $0.18 \pm 0.04$ \\
\citet{SantanaRos2025} & 2025 Jul 2--29 & \nodata & $0.42 \pm 0.02$ & $0.65 \pm 0.03$ & $0.27 \pm 0.03$ \\
\citet{Beniyama2025} & 2025 Jul 15 & \nodata & \nodata & $0.603 \pm 0.031$ & $0.210 \pm 0.031$ \\
\citet{SantanaRos20263I} & 2025 Jul 26--Sep 11 & $0.867 \pm 0.009$ & $0.542 \pm 0.005$ & $0.644 \pm 0.006$ & $0.265 \pm 0.005$ \\
\citet{Bolin2025DecATel} & 2025 Dec 22 & \nodata & \nodata & $0.69 \pm 0.02$ & $0.23 \pm 0.02$ \\
\citet{Bolin2026JanATel} & 2026 Jan 20--21 & \nodata & \nodata & $0.71 \pm 0.01$ & $0.25 \pm 0.02$ \\
\citet{Moreno26} & 2025 Jul 17--2026 Feb 18 & \nodata & \nodata & $0.70 \pm 0.02$ & $0.22 \pm 0.05$ \\
\enddata
\end{deluxetable*}

\section{Discussion}
\label{sec:discussion}

Our measured broadband colors show little evidence for systematic evolution over the observed portion of the apparition. Similar color stability was reported by \citet{Moreno26} across pre- and post-perihelion observations and by \citet{Gillan2026} and \citet{SantanaRos20263I} pre-perihelion. Stable coma colors have also been observed in the distantly active long-period comet C/2017 K2 (PANSTARRS), where they were attributed to relatively stable dust properties \citep{Hmiddouch2025}, and in active Main Belt asteroids \citep[e.g.,][]{Moreno2017J1,Ye2019Gault}. Our result is also consistent with the polarimetry of \citet{Choi2026}, who found no significant evolution in the $R_{\rm C}$-band polarization phase curve and interpreted this behavior as evidence for the sustained release of refractory dust with broadly similar optical and structural properties. We note that \citet{Tonry2025} found that the ATLAS $c-o$ color changed from red ($\sim0.7$) on 2025 July 4 to nearly solar ($\sim0.3$) by 2025 July 14. This evolution predates our first color measurement on 2025 July 17. Moreover, the broad ATLAS $c$ (420--650~nm) and $o$ (560--820~nm) bandpasses and different photometric apertures sample different combinations and spatial distributions of dust and gas, so the two datasets need not show identical color evolution.

The lack of detectable broadband color evolution is notable given the substantial evolution of the volatile coma described above. Several prominent molecular emission features fall within the optical bandpasses used here, including the CN violet bands near 388 and 422~nm, C$_3$ emission near 405~nm, and the C$_2$ Swan bands near 474, 517, and 564~nm. Because the strongest C$_2$ bands span both $B$ and $V$, changes in C$_2$ emission can affect both magnitudes and partially cancel in $B-V$. In contrast, the principal C$_2$ Swan bands fall largely within Sloan $g$ rather than $r$, so changes in C$_2$ emission may produce a stronger signature in $g-r$. We nevertheless find little evidence for systematic evolution in any of the measured color indices. Thus, despite substantial evolution of the gas coma, the combined contributions of molecular emission and dust scattering produced no detectable systematic evolution in the broadband colors measured within our projected $\sim$10{,}000~km apertures.

\section{Conclusions}
\label{sec:conclusions}

We presented a temporally dense, multi-facility optical color dataset of 3I, combining observations from the Unistellar Network with measurements obtained through a global relay of professional and citizen-operated telescopes. The observations span the inbound and outbound portions of the comet's 2025--2026 apparition and allow us to investigate its color behavior over a broad range of observing geometries and heliocentric distances. Our main conclusions are as follows:

\begin{enumerate}

\item Throughout the portion of the apparition sampled by our color measurements, the optical coma of 3I remained consistently red. We measure representative mean colors of
$B-V = 0.86 \pm 0.06$,
$V-R = 0.50 \pm 0.03$,
$B-R = 1.36 \pm 0.08$, and
$g-r = 0.58 \pm 0.08$.

\item We find little evidence for systematic long-term color evolution. Linear and piecewise-linear models do not provide a consistent improvement over the constant-color model, and the fitted trends differ among the four color indices. Although the comet underwent substantial changes in brightness and activity during its apparition, these changes were not accompanied by corresponding systematic changes in its optical color.

\item The observed color stability suggests that the ensemble-averaged optical scattering properties of the coma remained comparatively stable over the period sampled by our observations.

\end{enumerate}

\begin{acknowledgments}

This work was supported by the Gordon and Betty Moore Foundation and AstronetX.

This work is also supported by the Ministry of Education, Youth and Sports (MEYS, Czech Republic) under the projects MEYS LM2023032, LM2023047, and CZ.02.01.01/00/22\_008/0004632.

The Liverpool Telescope is operated on the island of La Palma by Liverpool John Moores University in the Spanish Observatorio del Roque de los Muchachos of the Instituto de Astrofisica de Canarias with financial support from the UK Science and Technology Facilities Council. Based on observations taken at the Liverpool Telescope during the PATAG proposal AZ22A01 (PI: F. Romanov).

Filipp Romanov is grateful to iTelescope.Net for providing him with complimentary observing time on their remote telescopes and to Daniel Parrott for providing the Tycho Tracker software free of charge. We thank T\"{U}B\.{I}TAK National Observatory for partial support in using the TUG100 telescope under the T\"{u}rkiye National Observatories with project number 25BTUG100-3023. We thank Adiyaman University Astrophysics Application and Research Center for partial support in using the ADYU60 telescope.

The National Research Institute of Astronomy and Geophysics team acknowledges financial support from the Egyptian Science, Technology \& Innovation Funding Authority (STDF) under grant number 48102. 

The research of the Astronomical Institute of the Slovak Academy of Sciences (MH, OI, DT) was supported by the Slovak Research and Development Agency under Contract No.~APVV-24-0076 and by the grant of the Slovak Academy of Sciences (VEGA No.~2/0067/26). This work was co-funded by the EU and supported by the Czech Ministry of Education, Youth and Sports through project CZ.02.01.01/00/22\_008/0004596 (SenDISo).

This research made use of the AAVSO Photometric All-Sky Survey (APASS), funded by the Robert Martin Ayers Sciences Fund and NSF AST-1412587. This work made use of Astropy, a community-developed core Python package and an ecosystem of tools and resources for astronomy.

This research benefited from B. T. Bolin's participation in the 2026 workshop ``Exploring Planetary Systems in the Era of Time-domain Astronomy,'' hosted by the Institute for Astronomy at the University of Hawai'i and supported by award 2106927 from the National Science Foundation's Astronomy \& Astrophysics Research Grants program.

O. A. Burkhonov and K. E. Ergashev are thankful for the state funding of the Department of Galactic Astronomy at the Astronomical Institute of the Academy of Sciences of Uzbekistan.

\end{acknowledgments}

\appendix

\section{Observing Facilities}

\label{app:facilities}

This appendix provides additional information on the telescopes, instrumentation, observing conditions, and data-reduction procedures where available for the facilities contributing observations to this work. Observatory codes refer to the Minor Planet Center (MPC) observatory-code system.

\subsection{Adiyaman University 0.60-m Telescope}

3I was monitored with the 0.60-m ADYU60 telescope\footnote{\url{https://observatory.adiyaman.edu.tr/tr/gozlem-aletleri/teleskoplar}} at the Adiyaman University Astrophysics Application and Research Center in Adiyaman, T\"urkiye. The ADYU60 is equipped with an Andor iKon-M 934 charge-coupled device (CCD) detector (1024$\times$1024 pixels). Observations were obtained on 2025 November 24 and December 14 using Johnson--Cousins $B$, $V$, $R$, and $I$ filters, comprising 213 frames in total. Seeing during the observations ranged from 3.0\arcsec--3.5\arcsec.


\subsection{GAL Hassin Observatory, MPC M57}

3I was imaged with the 1.0-m Wide-field Mufara Telescope (WMT) at GAL Hassin Observatory on Mt. Mufara, Sicily, Italy, on three nights: 2025 August 18 and 26 ($\sim$1 hr of observations on each night) and September 19 ($\sim$0.5 hr). The WMT is a 1.0-m-class prime-focus telescope built by Officina Stellare, with a focal ratio of $f/2.1$, and is equipped with a Spectral Instruments 1110s CCD camera. The telescope is located at an altitude of 1865 m \citep{Nastasi2023}. Images were obtained using Sloan $g$ and $r$ filters with the telescope tracking sidereally. Exposure times were kept sufficiently short to limit trailing of the comet in individual frames to less than 2 pixels. The image plate scale was 0.98\arcsec\,pixel$^{-1}$. Typical seeing during the observations was $\sim$2.2\arcsec--3.5\arcsec, and the airmass ranged from 2.0--4.7.

Images were calibrated using standard procedures, including bias and dark subtraction, flat-fielding, and alignment, with Tycho Tracker v.12\footnote{\url{https://www.tycho-tracker.com/}} \citep{Parrott2020}, which was also used for the photometric analysis. Photometry was extracted from the individual frames using apertures corresponding to a projected radius of 10{,}000~km at the comet. The signal-to-noise ratio (S/N) of the comet ranged from 20--70 in Sloan $g$ and from $\sim$30--120 in Sloan $r$. Differential photometry was calibrated using the ATLAS All-Sky Stellar Reference Catalog \citep{Tonry2018}, selecting comparison stars within 15 arcmin of the comet's mean position and with $0.5 \leq B-V \leq 1.0$.

\subsection{Cherenkov Telescope Array, MPC 950}

3I was imaged with the FRAM (F/Photometric Robotic Atmospheric Monitor) at the Cherenkov Telescope Array site on La Palma, Canary Islands, Spain, on 2025 November 15 \citep{Janecek2019FRAM}. The instrument uses a 135-mm focal-length telephoto lens with a focal ratio of $f/1.8$. Images were obtained using Johnson--Cousins $V$ and $R$ filters \citep{Cousins1976}. The instrument tracked sidereally, while comet-aligned stacking was performed during post-processing with Tycho Tracker \citep{Parrott2020}. Images were processed using an automated reduction pipeline including dark subtraction and flat-field correction, followed by image stacking and photometric measurements with Tycho Tracker. The airmass was 3.03 and the seeing was 2.1\arcsec. Because the angular resolution of FRAM is insufficient for reliable photometry within a projected 10{,}000~km radius, the photometric aperture was instead selected to encompass the entire visible coma.

\subsection{iTelescope.Net Deep Sky Chile, MPC X07}

3I was imaged with the 0.51-m T72 telescope, equipped with a KAF-16200 CCD sensor, and the 0.25-m T75 telescope, equipped with an IMX455 complementary metal-oxide-semiconductor (CMOS) sensor, at iTelescope.Net's Deep Sky Chile facility in the R\'io Hurtado Valley, Chile. T72 observations were obtained on three nights between 2025 September 11 and 27, while T75 observations were obtained on 2025 October 2. Images from T72 were obtained using Johnson--Cousins $B$, $V$, and $R_{\rm C}$ filters, while T75 observations used Chroma Red, Green, and Blue filters. Both telescopes tracked sidereally.

For T72, seeing during the observations ranged from 2.5\arcsec--3.0\arcsec, and the airmass ranged from 1.40--2.70. For T75, the seeing was approximately 7\arcsec and the airmass was 4.11. Images were processed using the automated reduction pipeline, and photometric measurements were performed with Tycho Tracker \citep{Parrott2020}.

\subsection{iTelescope.Net Sierra Remote Observatory, MPC U69}

3I was imaged with the 0.61-m T24 telescope at iTelescope.Net's Sierra Remote Observatory in Auberry, California, USA, on 2025 December 1. The telescope is equipped with an FLI-PL09000 CCD camera (3056$\times$3056 pixels). Images were obtained using a Johnson--Cousins $V$ filter and Astrodon Red and Blue filters with sidereal tracking. The seeing was 2.5\arcsec, and the airmass was 1.45. Images were processed using the automated reduction pipeline, and photometric measurements were performed with Tycho Tracker \citep{Parrott2020}.

\subsection{iTelescope.Net Utah Desert Remote Observatory, MPC U94}

3I was imaged with the 0.43-m T21 telescope, equipped with an FLI-PL6303E CCD, at iTelescope.Net's Utah Desert Remote Observatory in the Great Basin Desert near Beryl Junction, Utah, USA, on five nights between 2025 November 24 and December 28. Images were obtained using Johnson--Cousins $B$, $V$, and $R_{\rm C}$ filters with sidereal tracking. Seeing during the observations ranged from 2.5\arcsec--4.0\arcsec, and the airmass ranged from 1.12--1.67. Images were processed using the automated reduction pipeline, and photometric measurements were performed with Tycho Tracker \citep{Parrott2020}.

\subsection{Korea Microlensing Telescope Network, MPC L32}

3I was imaged with the Korea Microlensing Telescope Network (KMTNet) telescope at the South African Astronomical Observatory (SAAO), South Africa, on 15 nights between 2025 August 18 and September 10. Observations were obtained using Johnson--Cousins $B$, $V$, and $R$ filters. The airmass ranged from 1.04--1.56; seeing information is unavailable.

\subsection{Kottamia Astronomical Observatory 1.88-m Telescope, MPC 088}

3I was imaged with the 1.88-m telescope at Kottamia Astronomical Observatory (KAO), Egypt, using the Kottamia Faint Imaging SpectroPolarimeter \citep[KFISP,][]{Azzam2022} on nine nights between 2025 July 21 and December 23. Images were obtained using Johnson--Cousins $B$, $V$, $R$, and $I$ filters \citep{Cousins1976}, and the telescope was tracked at the comet's apparent sky-motion rate. The airmass ranged from 1.2--1.9 and the typical seeing during the observations was $\sim$1--3\arcsec. Bias subtraction and flat-fielding were applied to the images. 

\subsection{Kryoneri Observatory, MPC L10}

3I was imaged at Kryoneri Observatory in Greece on four nights between 2025 July 24 and 2026 February 2. Data were acquired using the 1.2~m Kryoneri telescope in its prime-focus configuration \citep{Xilouris2018}, equipped with an Andor Zyla 5.5 sCMOS camera. All images were obtained through a Cousins $I$ filter under sidereal tracking. The target airmass ranged from 1.7 to 3.2 during the July observations and from 1.2 to 1.7 during the February run. Standard reduction procedures were applied to all raw science frames, including bias subtraction, dark-frame correction, and flat-field calibration. Aperture photometry was performed relative to an ensemble of 3--5 field comparison stars calibrated against the \textit{Gaia} catalog.

\subsection{Liverpool Telescope, MPC J13}

3I was imaged with the 2.0-m Liverpool Telescope at the Roque de los Muchachos Observatory on La Palma, Canary Islands, Spain, using the IO:O camera \citep{Steele2004} on 15 nights between 2025 July 17 and 2026 February 21. Images were obtained using Sloan $g'$ and $r'$ filters. On some nights, the telescope was tracked at the comet's apparent sky-motion rate, while on others it tracked sidereally with sufficiently short exposures to prevent significant trailing of the comet. Typical seeing during the observations was 1\arcsec--2\arcsec, and the airmass ranged from 1.01--2.04. Images were processed using the automated reduction pipeline, and photometric measurements were performed with Tycho Tracker \citep{Parrott2020}.

\subsection{Los Angeles Astronomical Society Robotic Observatory}

3I was imaged with the Los Angeles Astronomical Society's Robotic Observatory on five nights between 2025 December 3 and 14. Images were obtained using Johnson--Cousins $B$, $V$, and $R$ filters. The airmass ranged from 1.21--1.54; seeing information is unavailable.

\subsection{Maidanak Observatory, MPC 188}

3I was imaged at Maidanak Observatory \citep{Ehgamberdiev2018} in Uzbekistan on 9 nights during August 2025 using the 1.5-m telescope equipped with a 4k $\times$ 4k CCD SNUCAM. Images were obtained using Bessell $BVR$ filters \citep{Im2010}.
Observations were conducted at airmasses between 1.8 and 3.2, with typical $R$-band seeing conditions varying from 1.5 to 2.5 arcseconds. The seeing quality in the $B$ and $V$ bands was 4\% and 8\% lower, respectively. See Table \ref{tab:geometry}, which lists the sequence for the $B$, $V$, and $R$ filters. Each epoch consisted of 10 images in each of the $B$, $V$, and $R$ filters, totaling 600 seconds of exposure time per filter. Photometry was performed on the stacked images, which were primarily aligned to compensate for the comet's motion.

\subsection{Observatori Astron\`omic de Les Planes de Son, MPC C29}

3I was imaged at the Observatori Astron\`omic de Les Planes de Son, Spain, on 2025 July 21 using the 20-inch CDK telescope with a focal length of 3454~mm, equipped with an Atik Horizon II CCD camera. Images were obtained using a Sloan $g$ filter.

\subsection{Observatori de Pujalt, MPC M04}

3I was imaged with the 0.50-m TGS telescope (Telescopi Guille i de Sol\`a) at the Observatori de Pujalt in Barcelona, Spain, on the night of 2025 August 20. Images were obtained using a Sloan $g'$ filter, and the telescope was tracked at the comet's apparent sky-motion rate. Flat-field calibration was performed to remove sensitivity variations and achieve a uniform image background. Observing conditions, including seeing and airmass, were not recorded for all observations and are therefore unavailable.

\subsection{Palomar Observatory, MPC 675}

3I was imaged with the 5.1-m (200-inch) Hale Telescope at Palomar Observatory in California, USA, on 2025 December 22. Images were obtained using Sloan $g$, $r$, and $i$ filters. The airmass ranged from 1.75--1.76; seeing information is unavailable.

\subsection{Pierre Auger Observatory, MPC I47}

3I/ATLAS was imaged with the 0.30-m FRAM (F/Photometric Robotic Atmospheric Monitor) telescope at the Pierre Auger Observatory between 2025 July 24 and 2026 February 18 \citep{aab2021}. Images were obtained on 15 nights using Johnson--Cousins $V$ and $R$ filters. The seeing during the observations ranged from 3.8\arcsec to 9.5\arcsec, and the airmass covered a range of 1.05––3.09. The telescope was tracked at the sidereal rate, while comet-aligned stacking was performed during post-processing in Tycho Tracker \citep{Parrott2020}. The images were processed using an automated reduction pipeline including dark subtraction, flat-field correction, and image stacking, followed by photometric measurements in Tycho Tracker. Because the angular resolution of FRAM is insufficient for reliable photometry within a projected 10{,}000~km radius, the photometric aperture was instead selected to encompass the entire visible coma.

\subsection{Schiaparelli Southern Observatory, MPC M21}

3I was imaged at the Schiaparelli Southern Observatory located at Hakos Farm, Namibia, on 7 nights between 2025 August 8 and 2025 September 23 using the Officina Stellare 0.36-m $f/8.6$ Ritchey--Chr\'etien ``Ferrante'' telescope equipped with a FLI FL16803 CCD until 2025 Aug. 16, then a QHY600M CMOS from 2025 Aug. 24. Images were obtained using Johnson--Cousins $BVR$ photometric filters with sidereal tracking. Single exposures were short enough that the comet was not trailed (less than 2 pixels), and images were stacked in order to obtain a good SNR. Seeing was not measured directly, but stellar FWHM values indicate typical seeing of approximately 2.5\arcsec--3.5\arcsec. All images were calibrated using standard procedures, including dark-frame subtraction and flat-field correction, with Astrometrica v.4.16.4.468. Photometry was performed using the same software with UCAC4 as the reference catalog.

\subsection{Skalnat\'e Pleso Observatory, MPC 056}

3I was imaged with the 0.61-m Newton telescope\footnote{\url{https://www.astro.sk/en/research/observatories/skalnate-pleso-observatory/}} at Skalnat\'e Pleso Observatory (SPO), Slovakia, on seven nights between 2025 July 3 and August 14. Images were obtained using broadband Johnson--Cousins $B$, $V$, and $R$ filters, and the telescope tracked sidereally. Typical seeing during the observations was $\sim$3.0\arcsec--4.0\arcsec, and the airmass ranged from 2.6--4.6. All images were calibrated using standard procedures, including dark-frame correction and flat-fielding. Photometry was obtained using aperture photometry relative to selected field stars, with calibration based on reference magnitudes from either the ATLAS All-Sky Stellar Reference Catalog \citep{Tonry2018} or the American Association of Variable Star Observers Photometric All-Sky Survey (APASS) catalog \citep{Henden2014}.

\subsection{Skygems Namibia, MPC L81}

3I was imaged at Skygems Namibia, Namibia, on 21 nights between 2025 July 7 and 2026 March 20. Images were obtained using green filters from luminance-red-green-blue (LRGB) filter sets and the telescope was tracked sidereally. Each observation involved multiple short exposures to keep motion blur of the comet below image resolution. Background star trails were subtracted on the combined (stacked) images where necessary. Photometric measurements were calibrated to Gaia EDR3 $G_\mathrm{BP}$ magnitudes, including a linear color transformation term.

\subsection{SkyMapper Network}

3I was imaged with telescopes in the SkyMapper Network, a globally distributed network of citizen-operated robotic reflecting telescopes, on 27 nights between 2025 November 23 and 2026 March 27. The telescopes used in this study were all Unistellar eVscopes. See Section \ref{sec:obs_inst_unistellar} for more information on the eVscope instruments. See Table \ref{tab:photometry} for the seeing information for each eVscope observation.

\subsection{South African Astronomical Observatory, MPC M28}

3I was imaged with the 1.0-m Lesedi telescope at the South African Astronomical Observatory (SAAO), South Africa, using the Mookodi instrument \citep{Erasmus2024}. The observations were obtained as part of the rapid follow-up program of the SAAO Intelligent Observatory Program \citep{Erasmus2025} on the nights of 2025 August 26 and September 1. Images were obtained using Sloan $g$, $r$, $i$, and $z$ filters. The telescope tracked sidereally, with exposure times limited to 30 s on both nights to avoid significant trailing due to the comet's apparent sky motion. Seeing during the observations was $\sim$1.2\arcsec and $\sim$1.3\arcsec on August 26 and September 1, respectively, with corresponding airmass ranges of 1.1--1.6 and 1.2--1.3. Bias subtraction and flat-fielding were applied to the images.

\subsection{Tivoli Astrofarm, MPC 194}

3I was imaged at Tivoli Astrofarm in Namibia on 6 nights between 2025 September 7 and 2025 November 10 (courtesy of M. J\"ager and G. Rhemann). Images were obtained using a green filter from an LRGB filter set and the telescope tracked sidereally. Each observation involved multiple short exposures to keep motion blur of the comet below image resolution. Background star trails were subtracted on the combined (stacked) images where necessary. Photometric measurements were calibrated to Gaia EDR3 $G_\mathrm{BP}$ magnitudes, including a linear color transformation term.

\subsection{T\"{U}B\.{I}TAK National Observatory, MPC A84}

3I was monitored with the 1.0-m TUG100 Ritchey--Chr\'etien telescope\footnote{\url{https://trgozlemevleri.gov.tr/teleskoplar/antalya/tug100}} at the T\"{U}B\.{I}TAK National Observatory (TUG) in Antalya, T\"urkiye. The TUG100 is equipped with an SI 1100 Cryo CCD detector (4096$\times$4037 pixels). Observations were obtained on 2025 November 20, December 18, and 2026 January 15 using Johnson--Cousins $B$, $V$, $R$, and $I$ filters, comprising 87 frames in total. The corresponding seeing values were 3.0\arcsec, 1.5\arcsec, and 1.5\arcsec, respectively.


\subsection{Unistellar Network, MPC 270}
\label{sec:obs_inst_unistellar}

3I was imaged with Unistellar eVscopes in a globally distributed network of citizen-operated robotic reflecting telescopes on 169 nights between 2025 July 3 and 2026 April 1. In total, the Unistellar dataset includes 405 observations contributed by 94 eVscopes.

The eVscopes contributing to this work are equipped with Sony CMOS image sensors arranged in red-green-blue (RGB) Bayer filter matrices. The three sensor models represented in our dataset are the IMX224, IMX347, and IMX415. The IMX224- and IMX347-based models, referred to hereafter as v1 and v2, respectively, have apertures of approximately 11.4 cm and focal ratios of $f/4$, corresponding to pixel scales of 1.72\arcsec\,pixel$^{-1}$ and 1.33\arcsec\,pixel$^{-1}$, respectively. Their fields of view are approximately 37\arcmin $\times$ 28\arcmin and 45.3\arcmin $\times$ 34\arcmin, respectively. The IMX415-based models, referred to hereafter as od (for the Odyssey model), have an aperture of 8.5 cm, a focal ratio of $f/3.9$, a pixel scale of 0.93\arcsec\,pixel$^{-1}$, and a field of view of approximately 45\arcmin $\times$ 34\arcmin.

Individual exposures were 4.0 s and were taken consecutively over observing sequences lasting 20--60 min. Because the Unistellar telescopes do not use standard astronomical filters, color information was extracted from the Bayer-filtered blue, green, and red channels and transformed to Johnson--Cousins $BVR$ as described in Section~\ref{sec:unistellar_color}. The telescopes tracked sidereally, so the science images were dark-subtracted and aligned to the comet's apparent motion before stacking. Flat-field frames were not required, as it was impractical to obtain them uniformly across the entire network. Observing conditions varied with time and observer location; therefore, the seeing and airmass for each observation are reported in Table~\ref{tab:geometry}.

\section{Observation Tables}
\label{app:observation_tables}

Table~\ref{tab:geometry} provides the observing geometry and observational circumstances for the measurements used in this work. Table~\ref{tab:photometry} provides the corresponding photometric measurements.

\startlongtable
\begin{deluxetable}{lllrrrrrrr}
\tabletypesize{\scriptsize}
\tablecaption{Observing geometry for 3I/ATLAS\label{tab:geometry}}
\tablewidth{0pt}
\tablehead{
  \colhead{Date and Time (UTC)}  &
  \colhead{Obs.\ Code\tablenotemark{a}} &
  \colhead{Location}             &
  \colhead{$r_H$\tablenotemark{b}} &
  \colhead{$\Delta$\tablenotemark{c}} &
  \colhead{$\alpha$\tablenotemark{d}} &
  \colhead{Airmass}              &
  \colhead{Seeing\tablenotemark{e}} &
  \colhead{$T - T_P$\tablenotemark{f}} \\
  \colhead{}                     &
  \colhead{}                     &
  \colhead{}                     &
  \colhead{(au)}                 &
  \colhead{(au)}                 &
  \colhead{(deg)}                &
  \colhead{}                     &
  \colhead{(arcsec)}             &
  \colhead{(days)}
}
\startdata
    2025 Jul 02 23:31:12 & L81 & Hakos, NA & 4.45 & 3.45 & 2.53 & 1.06 & \nodata & 118.5 \\
    2025 Jul 03 04:35:32 & 270 (v2) & Valley Center, CA, US & 4.44 & 3.44 & 2.60 & 2.21 & 8.51 & 118.3 \\
    2025 Jul 03 04:35:32 & 270 (v2) & Valley Center, CA, US & 4.44 & 3.44 & 2.60 & 2.21 & 8.51 & 118.3 \\
    2025 Jul 03 22:59:07 & 270 (v2) & Caen, FR & 4.42 & 3.42 & 2.84 & 2.65 & 7.12 & 117.5 \\
    2025 Jul 03 23:30:54 & 270 (v2) & Linhó, PT & 4.41 & 3.42 & 2.85 & 1.86 & 6.00 & 117.5 \\
    2025 Jul 04 00:28:48 & L81 & Hakos, NA & 4.41 & 3.42 & 2.86 & 1.20 & \nodata & 117.5 \\
    2025 Jul 04 03:44:56 & 270 (od) & Louisville, KY, US & 4.41 & 3.41 & 2.91 & 1.95 & 5.09 & 117.3 \\
    2025 Jul 04 06:02:59 & 270 (v2) & Cloverdale, CA, US & 4.41 & 3.41 & 2.94 & 2.02 & 7.98 & 117.2 \\
\enddata
\tablenotetext{a}{Minor Planet Center (MPC) observatory codes are given where available; otherwise, an abbreviation identifying the observing facility is used. For Unistellar observations (MPC code 270), the telescope version is additionally specified by detector: v1 corresponds to the IMX224 CMOS detector, v2 to the IMX347 CMOS detector, and od (Odyssey) to the IMX415 CMOS detector.}
\tablenotetext{b}{Heliocentric distance.}
\tablenotetext{c}{Geocentric distance.}
\tablenotetext{d}{Solar phase angle.}
\tablenotetext{e}{Seeing values are reported here only for Unistellar eVscope observations. Seeing measurements for other facilities, where available, are provided in Appendix~\ref{app:facilities}.}
\tablenotetext{f}{Time from perihelion, in days.}
\end{deluxetable}

\newpage

\begingroup

\setlength{\tabcolsep}{2pt}

\begin{longrotatetable}

\startlongtable

\begin{deluxetable}{lllrrrrrrrr}

\tabletypesize{\scriptsize}

\tablecaption{Photometric observations of 3I/ATLAS.\label{tab:photometry}}

\tablewidth{0pt}
\tablehead{
  \colhead{Date and Time (UTC)} & \colhead{Obs.\ Code\tablenotemark{a}} & \colhead{Location} &
  \colhead{Ap.\ Rad.} & \colhead{$m_G$\tablenotemark{b}} &
\colhead{$B$\tablenotemark{c}} &
\colhead{$V$\tablenotemark{c}} &
\colhead{$R$\tablenotemark{c}} & 
\colhead{$g$\tablenotemark{d}} &
\colhead{$r$\tablenotemark{d}} &
\colhead{$i$\tablenotemark{d}} \\
  \colhead{} & \colhead{} & \colhead{} & \colhead{(arcsec)} & \colhead{(mag)} &
  \colhead{(mag)} & \colhead{(mag)} & \colhead{(mag)} & \colhead{(mag)} &
  \colhead{(mag)} & \colhead{(mag)}}
\startdata
    2025 Jul 02 23:31:12 & L81 & Hakos, NA & 4.00 & \nodata & \nodata & $18.49 \pm 0.07$ & \nodata & \nodata & \nodata & \nodata \\
    2025 Jul 03 04:35:32 & 270 (v2) & Valley Center, CA, US & 4.01 & $17.73 \pm 1.06$ & \nodata & \nodata & \nodata & \nodata & \nodata & \nodata \\
    2025 Jul 03 04:35:32 & 270 (v2) & Valley Center, CA, US & 4.01 & $17.61 \pm 0.45$ & \nodata & \nodata & \nodata & \nodata & \nodata & \nodata \\
    2025 Jul 03 22:59:07 & 270 (v2) & Caen, FR & 4.03 & $18.99 \pm 1.67$ & \nodata & \nodata & \nodata & \nodata & \nodata & \nodata \\
    2025 Jul 03 23:30:54 & 270 (v2) & Linhó, PT & 4.04 & $16.25 \pm 0.16$ & \nodata & \nodata & \nodata & \nodata & \nodata & \nodata \\
    2025 Jul 04 00:28:48 & L81 & Hakos, NA & 4.03 & \nodata & \nodata & $18.36 \pm 0.07$ & \nodata & \nodata & \nodata & \nodata \\
    2025 Jul 04 03:44:56 & 270 (od) & Louisville, KY, US & 4.04 & $17.13 \pm 0.33$ & \nodata & \nodata & \nodata & \nodata & \nodata & \nodata \\
    2025 Jul 04 06:02:59 & 270 (v2) & Cloverdale, CA, US & 4.04 & $17.62 \pm 1.24$ & \nodata & \nodata & \nodata & \nodata & \nodata & \nodata \\
\enddata
\tablenotetext{a}{Minor Planet Center (MPC) observatory codes are given where available; otherwise, an abbreviation identifying the observing facility is used. For Unistellar observations (MPC code 270), the telescope version is additionally specified by detector: v1 corresponds to the IMX224 CMOS detector, v2 to the IMX347 CMOS detector, and od (Odyssey) to the IMX415 CMOS detector.}

\tablenotetext{b}{$m_G$ denotes photometry calibrated to the Gaia $G$ band.}

\tablenotetext{c}{$B$, $V$, and $R$ denote Johnson--Cousins photometry.}

\tablenotetext{d}{$g$, $r$, and $i$ denote Sloan photometry.}

\end{deluxetable}

\end{longrotatetable}

\endgroup

\bibliography{bib.bib}
\bibliographystyle{aasjournal}

\end{document}